\documentclass[journal]{IEEEtran} 
\IEEEoverridecommandlockouts
\usepackage{cite}
\usepackage{amsmath,amssymb,amsfonts}
\usepackage{algorithmic}
\usepackage{graphicx}
\usepackage[numbers]{natbib}
\usepackage[export]{adjustbox}
\usepackage{textcomp}
\usepackage{xcolor}
\def\BibTeX{{\rm B\kern-.05em{\sc i\kern-.025em b}\kern-.08em
    T\kern-.1667em\lower.7ex\hbox{E}\kern-.125emX}}
\begin{document}

\title{H\-DNA\\}

\author{
    \IEEEauthorblockN{1\textsuperscript{st} Mahdi Akhi}
    , \IEEEauthorblockN{2\textsuperscript{nd} Nona Ghazizadeh}\\ 
    \IEEEauthorblockA{\textit{dept. Computer Engineering (of Sharif University of Technology), Tehran, Iran}}\\
    \IEEEauthorblockA{\textit{dept. Computer Engineering (of Sharif University of Technology), Tehran, Iran}}\\
    \IEEEauthorblockA{mahdiakhi@sharif.edu, nona.ghazizadeh@sharif.edu}
}

\maketitle

\begin{abstract}
In this paper a new approach called HDNA (HTML DNA) is introduced for analyzing and comparing Document Object Model (DOM) trees in order to detect differences in HTML pages. This method assigns an identifier to each HTML page based on its structure, which proves to be particularly useful for detecting variations caused by server side updates, user interactions or potential security risks. The process involves preprocessing the HTML content generating a DOM tree and calculating the disparities between two or more trees. By assigning weights to the nodes valuable insights about their hierarchical importance are obtained. The effectiveness of the HDNA approach has been demonstrated in identifying changes in DOM trees even when dynamically generated content is involved. Not does this method benefit web developers, testers and security analysts by offering a deeper understanding of how web pages evolve. It also helps ensure the functionality and performance of web applications. Additionally it enables detection and response to vulnerabilities that may arise from modifications in DOM structures. As the web ecosystem continues to evolve HDNA proves to be a tool, for individuals engaged in web development, testing or security analysis.

\end{abstract}

\section{Introduction}
With the widespread adoption of the internet, HTML has emerged as the de-facto standard for designing and structuring web pages. The Document Object Model (DOM) is a platform-independent and language-independent representation of an HTML page that enables developers to interact with web content programmatically. \cite{DBLP:conf/www/ChakrabartiM10} \cite{DBLP:journals/corr/abs-2101-02415} It serves as a foundation for numerous web-related tasks, such as rendering, event handling, and dynamic updates. As web applications become increasingly complex and dynamic, the ability to identify and understand differences in DOM tree structures becomes vital. For instance, Progressive Web App (PWA) technologies such as ReactJS utilize DOM modifications to selectively re-render elements. This allows ReactJS to update only the altered element rather than the entire page. In fact, ReactJS responds to changes in elements, hence its name. By detecting changes within the PWA framework, web pages can be made faster and more lightweight. \cite{10.1007/978-3-319-93527-0_4} Changes in DOM trees can result from various factors, including server-side updates, user interactions, or injected malicious code. Accurately detecting and analyzing these structural differences in the DOM trees is essential for web developers, testers, and security analysts alike. Understanding these changes can help improve website performance, enhance user experience, and increase security. \cite{10.1007/978-3-319-93527-0_4}
In this paper, we present a novel approach for calculating and comparing DOM trees to identify differences in HTML pages. Each HTML page has a unique structure and arrangement of tags. By leveraging this characteristic, we can assign a unique identifier to each page that represents its structure. We refer to this unique identifier as \textbf{\textit{HDNA}} which stands for HTML DNA. We use the HDNA for each page to detect the changes or difference between two or more HTML pages. However, this approach can be used for any tree structure. Our approach leverages advanced algorithms and techniques to efficiently capture structural changes in DOM trees, even in the presence of dynamically generated content. By analyzing the hierarchical relationships, node attributes, and content variations, our method provides a detailed and accurate representation of the differences between DOM trees.

\section{Use case}
\textbf{HDNA} has a wide range of applications in web page technologies. It can be used to identify web pages, tags, or sub-trees. \cite{8333421} Additionally, HDNA can assist in detecting differences and changes between two or more pages or other tree-structured documents, such as XML. In the following, we mention some of the things that can be helped by HDNA.

Web application testing plays a crucial role in ensuring the functionality, reliability, and security of web-based systems. \cite{Petukhov2008DetectingSV} As web applications evolve and undergo frequent updates, it becomes essential to accurately detect and analyze structural differences in HTML pages to ensure that desired changes are implemented correctly and unintended side effects are avoided. HDNA involves calculating and comparing DOM trees to identify differences in HTML pages, can be applied to enhance various aspects of web application testing and also more potential use cases:

\subsection{Regression Testing}
Throughout the dynamic development lifecycle of web applications, a spectrum of alterations is introduced, spanning from minor bug fixes to substantial feature enhancements. Our methodology proves to be invaluable in facilitating the process of regression testing. \cite{Soto-Sánchez2022} It achieves this by adeptly automating the identification and conspicuous highlighting of disparities residing within the DOM (Document Object Model) trees of the freshly updated HTML pages. This pivotal functionality enables testers to strategically allocate their efforts towards scrutinizing the specific regions impacted by the changes. This meticulous focus ensures a thorough evaluation, assuring that the introduced modifications have not inadvertently introduced any unexpected repercussions or adverse effects into the system. This proactive approach is instrumental in maintaining the integrity, functionality, and reliability of web applications as they undergo iterative enhancements and refinements. Additionally, it empowers development teams to streamline their testing efforts and expedite the release of bug-free updates. \cite{10.1145/3442694}

\subsection{Automated Testing and Test Oracles}
Automated testing frameworks are heavily dependent on dependable test oracles, which serve as blueprints for the anticipated behavior of an application. \cite{8004387} Through the adoption of our methodology, which involves the meticulous comparison of DOM (Document Object Model) trees between reference and tested HTML pages \cite{DBLP:journals/corr/abs-2101-02415}, automated testing systems gain the ability to systematically assess the accuracy of the application's outputs. When differences are identified during this process, they act as red flags, signaling the potential presence of bugs or regressions in the application's functionality. This automated bug detection mechanism is an instrumental facet of enhancing the efficiency of testing procedures \cite{NooraeiAbadeh2021}, significantly diminishing the need for manual intervention in the creation and execution of test cases. This streamlined approach not only expedites the testing phase but also enhances the overall quality and reliability of the software being developed. Consequently, it fosters more robust and resilient applications in an increasingly dynamic and competitive digital landscape. \cite{Pai2021}

\subsection{Security Testing}
Web applications frequently find themselves in the crosshairs of malicious attackers, who employ tactics such as code injection or manipulation of the Document Object Model (DOM) structure to exploit vulnerabilities. \cite{Alkhalil2021PhishingAA} \cite{Humayun2020} In light of this persistent threat landscape, your chosen methodology stands as a formidable defense. It undertakes the crucial task of scrutinizing and contrasting DOM trees, effectively serving as a vigilant sentinel against unauthorized modifications or surreptitious code injections. This multifaceted capability creates a robust mechanism for preemptively identifying potential security breaches, fortifying the web application's defenses against cyber threats.
\\
The significance of this approach extends beyond mere detection; it revolutionizes the landscape of security testing. \cite{10.1007/978-3-319-73721-8_6} By automating the identification of malevolent alterations and assisting in vulnerability assessment, it not only bolsters the security posture of web applications but also significantly streamlines the security testing process. \cite{9278437} This proactive stance is invaluable in an era where cybersecurity threats continue to evolve in complexity and frequency. Furthermore, it underscores the commitment to safeguarding sensitive data, user privacy, and the overall integrity of web applications in an interconnected digital ecosystem. \cite{7494156} \cite{9143018} \cite{8342469}

\subsection{Deface attack detection}
Leveraging the capabilities of HDNA, which excels at computing disparities within DOM trees, we can employ it as a powerful tool for detecting deface attacks. \cite{electronics12122664} This entails a systematic process: at regular intervals, we retrieve the HTML content of a web page and meticulously calculate the contrast between the current content and the previously recorded content. If this calculated difference surpasses a predefined threshold that aligns with our expectations, we flag this alteration as a potential deface attack.
\\
This proactive strategy not only capitalizes on the precision of HDNA but also augments the security posture of web applications. By instituting automated, periodic checks for changes in HTML content, we establish a robust early-warning system for defacement attempts. \cite{Cremer2022} Furthermore, this approach promotes the rapid identification and mitigation of security breaches, safeguarding the integrity and reliability of web resources. Beyond the prevention of defacement, this methodology can also be extended to fortify web applications against a spectrum of malicious activities, including unauthorized alterations and data tampering \cite{10.1007/978-3-319-73721-8_6}, fostering a more secure online environment. \cite{Hehir2021}

\subsection{Web development}
In addition to its application in web application testing, our approach offers invaluable utility for web developers seeking to enhance the quality and functionality of their web applications. \cite{9673464} It serves as a versatile tool that empowers developers to efficiently uncover and troubleshoot issues. Through the precise identification and thorough analysis of structural variances within DOM trees, developers can swiftly pinpoint the root causes of problems. This proactive identification of issues accelerates the debugging process, facilitating the implementation of necessary adjustments and refinements. Consequently, web applications can be fine-tuned and improved, resulting in enhanced user experiences and more robust, error-free systems. \cite{9673464}
\\
Furthermore, the utilization of our approach extends beyond just debugging. It also promotes a more streamlined and collaborative development environment. \cite{10.1007/978-3-030-78221-4_21} Developers can utilize it as a means to comprehensively understand the intricate interplay of elements within their web applications, fostering a deeper insight into their code base. This holistic perspective can lead to more efficient development cycles, shorter time-to-market, and a competitive edge in the ever-evolving digital landscape. \cite{9673464}

\section{HDNA} 
\begin{figure}[h]
    \centering
    \includegraphics[width=\linewidth]{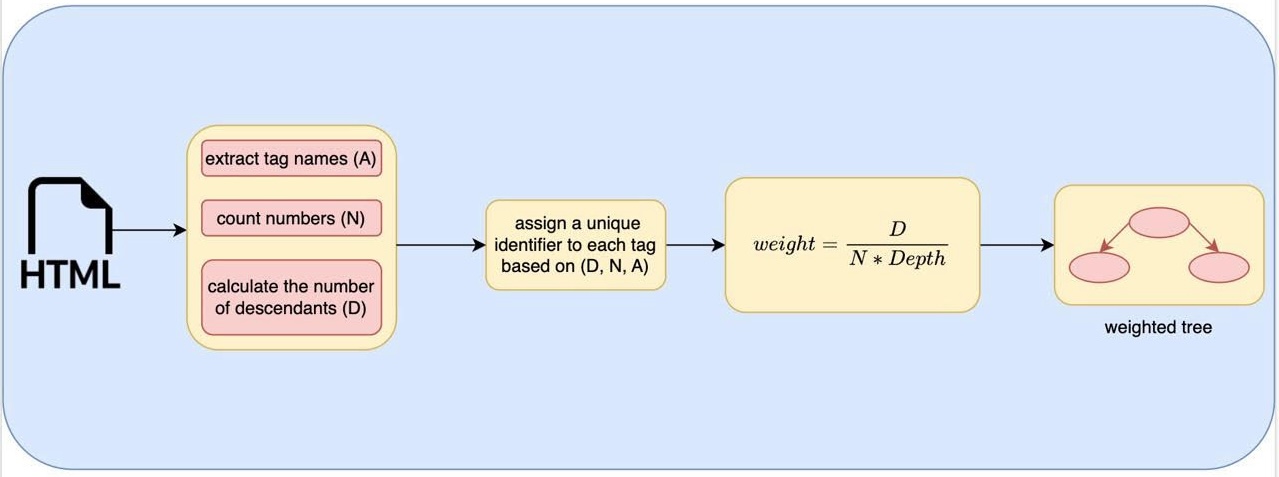}
    \caption{The overall pipelines and Model Architecture are shown here.}
    \label{fig:flow}
\end{figure}

The HDNA consist of two parts. At the first part, we introduce the $DNA$ which is a unique ID for nodes in DOM tree. Then at the second part, we describe how can we calculate the difference using $DNA$. Observe the flow as depicted in the Fig.\ref{fig:flow}.

\subsection{DNA}
In this section, we will introduce HDNA. As mentioned in the introduction, each HTML page has a unique structure. In HDNA, we assign a unique identifier to each tag based on its properties. We consider three properties to create a unique ID for the tags.

\begin{figure*}[h]
\centering
  \includegraphics[width=\textwidth, frame]{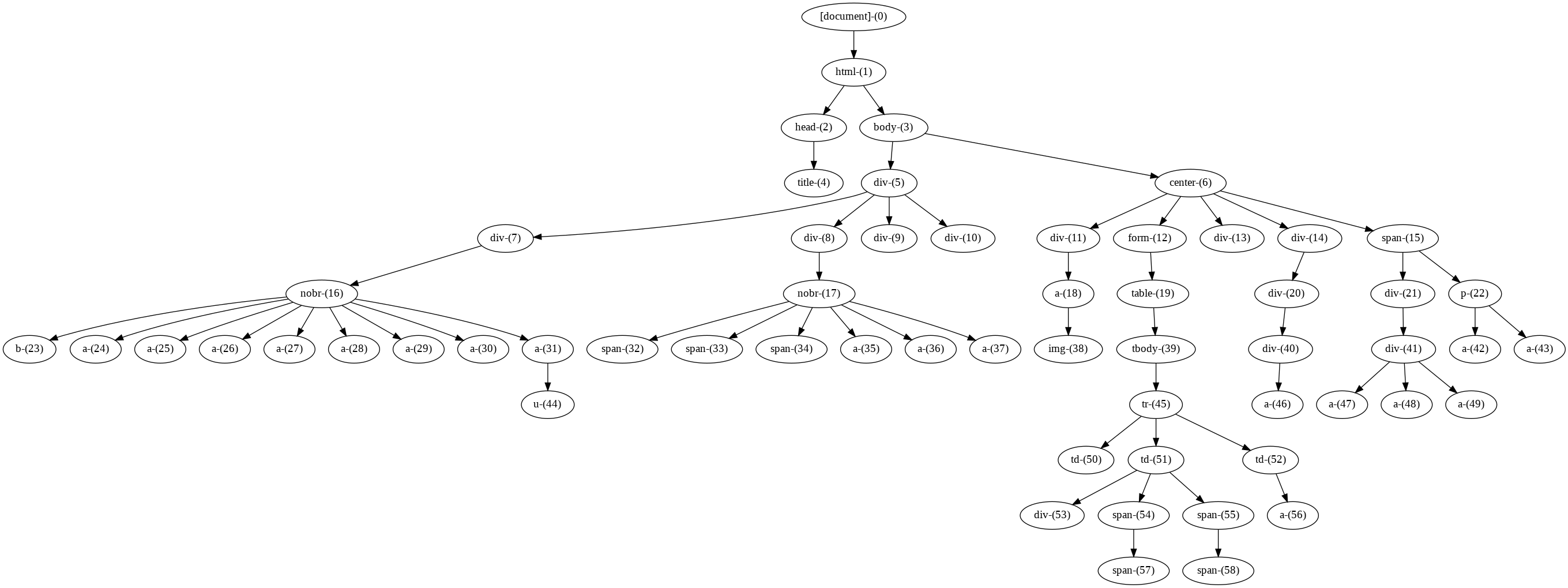} 
  \caption{Example of count number for Google first page}
  \label{fig:countNumberTree}
\end{figure*}

The first property considered in HDNA is the tag name, denoted by $A$. Each tag has a name, such as div, span, a, p, head, body, etc.
The second property is the count number, denoted by $N$. If we represent the DOM as a tree, the root node is the document and the other nodes are its children. If we assign a count number of zero to the document node, then the other nodes are numbered incrementally from left to right at each level, similar to a complete binary tree. Figure~\ref{fig:countNumberTree} illustrates this numbering for the Google homepage.
The third property is the number of descendants, denoted by $D$, which refers to the number of nodes in a tag’s sub-tree. Thus, we can assign an ordered triple $(D,N,A)$ to each tag (DOM node), which is unique for each node and allows us to compute difference.

\subsection{Difference Calculation}
To calculate differences between multiple trees, we need to designed a weighting mechanism. Each tag has a weight which calculates by equation~\ref{eq:weight}.

\begin{equation}\label{eq:weight}
\begin{split}
& Weight = \frac{D}{N*Depth}
\end{split}
\end{equation}

\begin{figure*}[h]
\centering
  \includegraphics[width=\textwidth, frame]{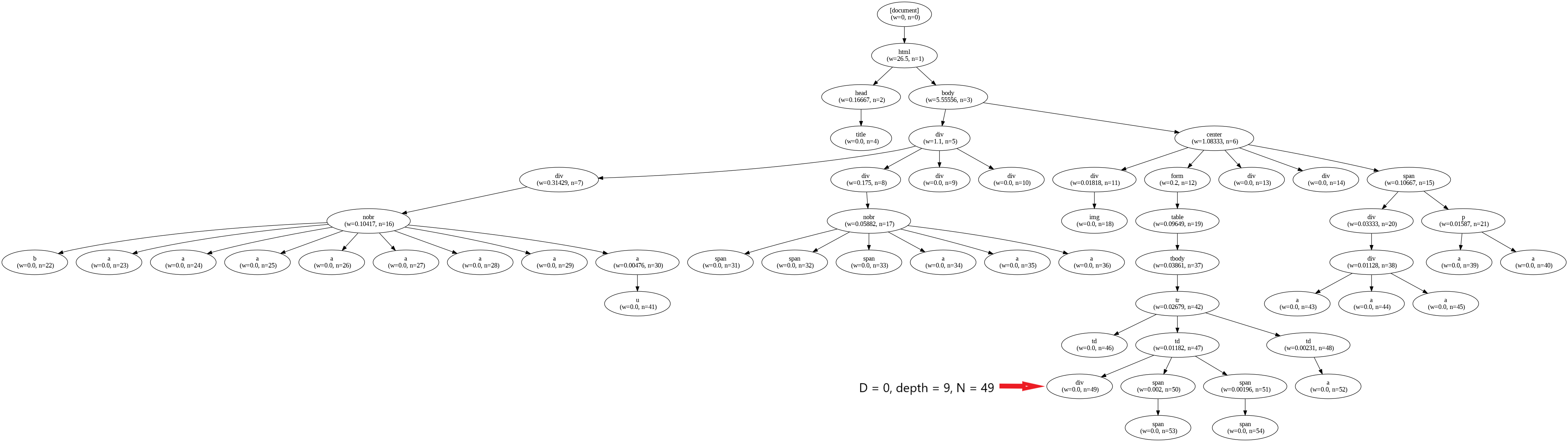} 
  \caption{The weighted nodes of DOM tree for the Google first page}
  \label{fig:HDNA}
\end{figure*}

The weight of each tag, as calculated by equation~\ref{eq:weight}, represents its importance. The weight is directly proportional to the number of descendants and inversely proportional to both the count number and depth of the tag. In other words, a tag that is located higher in the DOM tree and has a large sub-tree is considered more important. If such a tag is changed or removed, it can result in significant changes to the DOM tree. In the fig.~\ref{fig:HDNA} the weighted nodes for the Google first page is illustrated and the values for \ref{eq:weight} has shown for Node n=49. 

After calculate the weights of all nodes we can calculate the difference between trees. When you have the $DNA$ and the $weight$ for the trees, you are able to calculate the difference in different ways and it is depend on your problem. For example if you want to use HDNA to detect the deface attacks you can chain all tags' DNA in a string and get compare them to find whether the HTML has changed, Then if the HTML changed you can traverse the old and new HTML tree using the DNAs and when find a difference use its weight to realize the difference weight. 

\section{Implementation}
Implementation of this approach consists of several sections including preprocessing, generating DOM tree, weighting, and calculating the difference. All the codes are available in GitHub repository\footnote{https://github.com/mahdiAkhi/HDNA}
\subsection{Preprocessing}

When we initiate the retrieval of a web page using the \textit{request} library, we subsequently employ BeautifulSoup with the lxml parser to extract the BeautifulSoup object. Within this framework, we offer a range of specialized functions designed to enhance the utility of this object. One such function facilitates the selective removal of specific HTML tags from the BeautifulSoup object. This functionality proves invaluable for filtering out tags that are deemed extraneous or irrelevant for subsequent processing or analysis. Notably, the list of tags subject to removal includes: $script, meta, br, hr, link, input, style$.
\\
Another noteworthy function within our toolkit targets the removal of all attributes associated with HTML tags present in a BeautifulSoup object. It identifies tags bearing attributes, systematically eliminates those attributes, and subsequently delivers the modified BeautifulSoup object. This capability streamlines the structure and content of the object, often proving essential in data extraction and analysis tasks.
\\
Additionally, we provide a function that accomplishes a nuanced task: the extraction of textual content from an HTML tag and its nested children while preserving the inherent tag structure. This function is particularly advantageous when the objective is to retain the hierarchical arrangement of tags while removing the textual content itself. Such a feature facilitates subsequent processing or in-depth analysis, contributing to a more sophisticated and granular understanding of the web page's content and structure. These capabilities collectively empower users to wield BeautifulSoup effectively, making it a versatile asset for web scraping and data extraction endeavors.

\subsection{Generate Dom tree and weighting}
Following the preprocessing phase, we embark on a recursive journey through the BeautifulSoup object, employing the Python-based \textit{AnyTree} library to construct a hierarchical, tree-like structure. In this newly created structure, each HTML tag finds its counterpart as a distinct node within the tree. The result of this process is a graphical representation of the tree, offering a visual perspective that holds significant utility for subsequent stages of HTML document processing and analysis.
\\
We take a step further by ascribing weights to the nodes within this tree structure. These weights are derived from a combination of factors, including the number of descendants, the depth of each node within the tree, and its specific position as denoted by the node number. This weighting system imparts valuable information about the structural and hierarchical significance of each node. It serves as a critical dimension for assessing the importance and role of different elements within the HTML document.
\\
In essence, our approach not only transforms the HTML content into a comprehensible tree-like structure but also enriches it with contextual information in the form of node weights. This combination of structural representation and node weighting equips researchers, analysts, and developers with a powerful tool set for in-depth analysis, pattern recognition, and data-driven decision-making when dealing with complex HTML documents.

\subsection{Calculate difference}
We provide a function that takes two tree structures as input and compares the weights and names of corresponding nodes in the two trees. It returns a list of nodes that have differences between the trees. This function can be used to identify and analyze changes between two tree structures for our purpose.

\section{Case Studies}
Our method operates independently of data. Nevertheless, we tested it on 50 well-known websites covering a diverse range of subjects, including News, Sports, Media, and E-Commerce. We computed the HDNA for each website and arranged the results in DOT files, preserving the DOM tree structures within the files. All the data and outputs are available on the Kaggle\footnote{https://www.kaggle.com/datasets/mahdiakhi/hdna-output} and the Zenodo. 

\section{Future works}

Beyond the current implementation of DOM tree generation from HTML pages, there exist multiple promising avenues for both exploration and enhancement. Firstly, the handling of intricate web pages that host dynamically generated content could benefit from refinement in parsing logic and the integration of advanced parsing techniques. By doing so, we can better accommodate the intricacies of modern web content and improve the accuracy and completeness of the generated DOM tree. 
\\
Expanding the scope of our code to encompass the parsing and DOM tree generation of markup languages beyond HTML, such as XML or XHTML, represents another exciting possibility. This extension would substantially boost the versatility of the tool, making it applicable in a broader array of scenarios and content types. \cite{Choi2016}
\\
Moreover, considering the integration of DOM manipulation capabilities is a forward-looking step. \cite{DBLP:journals/corr/RabinovichSK17} By empowering users to execute operations like element addition, removal, or modification within the generated DOM tree, we furnish a comprehensive toolkit that serves not only web development but also content management purposes. This functionality holds the potential to streamline web page editing and content updates, making it a valuable asset for webmasters and content creators.
\\
Furthermore, there's room for exploration in the realm of performance optimization. Investigating techniques to enhance the efficiency of DOM tree generation, especially when dealing with large-scale or resource-intensive web pages, can yield significant benefits. This could involve strategies to expedite the parsing process or to handle memory-intensive operations more effectively, ultimately leading to a smoother and more efficient user experience. \cite{DBLP:journals/corr/abs-2009-10924} In conclusion, the potential for refinement and expansion in our DOM tree generation tool is abundant, promising a more versatile, efficient, and user-friendly solution for web developers and content managers alike.

\section{Conclusion}
In conclusion, our novel approach, HDNA, presents a powerful and efficient method for calculating and comparing DOM trees to identify differences in HTML pages. As the web continues to evolve, with increasingly complex and dynamic applications, the ability to accurately detect and analyze structural changes in DOM trees has become more vital than ever before. Our method, which assigns a unique identifier to each HTML page based on its structure, has demonstrated its effectiveness in detecting variations caused by server-side updates, user interactions, or potential security threats.
\\
Our approach not only benefits web developers by providing a deeper understanding of how web pages evolve but also aids testers in ensuring the functionality and performance of web applications. Moreover, security analysts can use our method to quickly identify and respond to potential vulnerabilities introduced by changes in DOM structures.
\\
By leveraging advanced algorithms and techniques, we have shown that our method can efficiently capture structural changes in DOM trees, even in the presence of dynamically generated content. This level of accuracy and granularity in detecting differences can lead to significant improvements in website performance, enhanced user experiences, and strengthened security measures.
\\
As the web ecosystem continues to evolve, we believe that the HDNA approach can be a valuable tool for anyone involved in web development, testing, or security analysis. Its versatility and robustness make it suitable for a wide range of applications beyond HTML, allowing for the analysis of various tree structures. We look forward to further research and development in this field, aiming to continually enhance and refine our approach to meet the evolving challenges of the web landscape. Ultimately, HDNA empowers web professionals with the insights and tools needed to build, test, and secure the web applications of tomorrow.

\bibliography{main-reference}

\begin{thebibliography}{24}
\providecommand{\natexlab}[1]{#1}
\providecommand{\url}[1]{#1}
\csname url@samestyle\endcsname
\providecommand{\newblock}{\relax}
\providecommand{\bibinfo}[2]{#2}
\providecommand{\BIBentrySTDinterwordspacing}{\spaceskip=0pt\relax}
\providecommand{\BIBentryALTinterwordstretchfactor}{4}
\providecommand{\BIBentryALTinterwordspacing}{\spaceskip=\fontdimen2\font plus
\BIBentryALTinterwordstretchfactor\fontdimen3\font minus
  \fontdimen4\font\relax}
\providecommand{\BIBforeignlanguage}[2]{{%
\expandafter\ifx\csname l@#1\endcsname\relax
\typeout{** WARNING: IEEEtranN.bst: No hyphenation pattern has been}%
\typeout{** loaded for the language `#1'. Using the pattern for}%
\typeout{** the default language instead.}%
\else
\language=\csname l@#1\endcsname
\fi
#2}}
\providecommand{\BIBdecl}{\relax}
\BIBdecl

\bibitem[Chakrabarti and Mehta(2010)]{DBLP:conf/www/ChakrabartiM10}
\BIBentryALTinterwordspacing
D.~Chakrabarti and R.~R. Mehta, ``The paths more taken: matching {DOM} trees to
  search logs for accurate webpage clustering,'' in \emph{Proceedings of the
  19th International Conference on World Wide Web, {WWW} 2010, Raleigh, North
  Carolina, USA, April 26-30, 2010}, M.~Rappa, P.~Jones, J.~Freire, and
  S.~Chakrabarti, Eds.\hskip 1em plus 0.5em minus 0.4em\relax {ACM}, 2010, pp.
  211--220. [Online]. Available: \url{https://doi.org/10.1145/1772690.1772713}
\BIBentrySTDinterwordspacing

\bibitem[Zhou et~al.(2021)Zhou, Sheng, Vo, Edmonds, and
  Tata]{DBLP:journals/corr/abs-2101-02415}
\BIBentryALTinterwordspacing
Y.~Zhou, Y.~Sheng, N.~Vo, N.~Edmonds, and S.~Tata, ``Simplified {DOM} trees for
  transferable attribute extraction from the web,'' \emph{CoRR}, vol.
  abs/2101.02415, 2021. [Online]. Available:
  \url{https://arxiv.org/abs/2101.02415}
\BIBentrySTDinterwordspacing

\bibitem[Bi{\o}rn-Hansen et~al.(2018)Bi{\o}rn-Hansen, Majchrzak, and
  Gr{\o}nli]{10.1007/978-3-319-93527-0_4}
A.~Bi{\o}rn-Hansen, T.~A. Majchrzak, and T.-M. Gr{\o}nli, ``Progressive web
  apps for the unified development of mobile applications,'' in \emph{Web
  Information Systems and Technologies}, T.~A. Majchrzak, P.~Traverso, K.-H.
  Krempels, and V.~Monfort, Eds.\hskip 1em plus 0.5em minus 0.4em\relax Cham:
  Springer International Publishing, 2018, pp. 64--86.

\bibitem[Petukhov and Kozlov(2008)]{Petukhov2008DetectingSV}
\BIBentryALTinterwordspacing
A.~Petukhov and D.~D. Kozlov, ``Detecting security vulnerabilities in web
  applications using dynamic analysis with penetration testing,'' 2008.
  [Online]. Available: \url{https://api.semanticscholar.org/CorpusID:1063530}
\BIBentrySTDinterwordspacing

\bibitem[Soto-S{\'a}nchez et~al.(2022)Soto-S{\'a}nchez, Maes-Bermejo, Gallego,
  and Gort{\'a}zar]{Soto-Sánchez2022}
\BIBentryALTinterwordspacing
{\'O}.~Soto-S{\'a}nchez, M.~Maes-Bermejo, M.~Gallego, and F.~Gort{\'a}zar, ``A
  dataset of regressions in web applications detected by end-to-end tests,''
  \emph{Software Quality Journal}, vol.~30, no.~2, pp. 425--454, Jun 2022.
  [Online]. Available: \url{https://doi.org/10.1007/s11219-021-09566-x}
\BIBentrySTDinterwordspacing

\bibitem[Bluemke and Malanowska(2021)]{10.1145/3442694}
\BIBentryALTinterwordspacing
I.~Bluemke and A.~Malanowska, ``Software testing effort estimation and related
  problems: A systematic literature review,'' \emph{ACM Comput. Surv.},
  vol.~54, no.~3, apr 2021. [Online]. Available:
  \url{https://doi.org/10.1145/3442694}
\BIBentrySTDinterwordspacing

\bibitem[Klammer and Ramler(2017)]{8004387}
C.~Klammer and R.~Ramler, ``A journey from manual testing to automated test
  generation in an industry project,'' in \emph{2017 IEEE International
  Conference on Software Quality, Reliability and Security Companion (QRS-C)},
  2017, pp. 591--592.

\bibitem[Nooraei~Abadeh(2021)]{NooraeiAbadeh2021}
\BIBentryALTinterwordspacing
M.~Nooraei~Abadeh, ``Genetic-based web regression testing: an ontology-based
  multi-objective evolutionary framework to auto-regression testing of web
  applications,'' \emph{Service Oriented Computing and Applications}, vol.~15,
  no.~1, pp. 55--74, Mar 2021. [Online]. Available:
  \url{https://doi.org/10.1007/s11761-020-00312-y}
\BIBentrySTDinterwordspacing

\bibitem[Pai et~al.(2021)Pai, Joshi, and Rane]{Pai2021}
\BIBentryALTinterwordspacing
A.~R. Pai, G.~Joshi, and S.~Rane, ``Quality and reliability studies in software
  defect management: a literature review,'' \emph{International Journal of
  Quality {\&} Reliability Management}, vol.~38, no.~10, pp. 2007--2033, Jan
  2021. [Online]. Available: \url{https://doi.org/10.1108/IJQRM-07-2019-0235}
\BIBentrySTDinterwordspacing

\bibitem[Alkhalil et~al.(2021)Alkhalil, Hewage, Nawaf, and
  Khan]{Alkhalil2021PhishingAA}
\BIBentryALTinterwordspacing
Z.~Alkhalil, C.~Hewage, L.~F. Nawaf, and I.~A. Khan, ``Phishing attacks: A
  recent comprehensive study and a new anatomy,'' in \emph{Frontiers of
  Computer Science}, 2021. [Online]. Available:
  \url{https://api.semanticscholar.org/CorpusID:232144884}
\BIBentrySTDinterwordspacing

\bibitem[Humayun et~al.(2020)Humayun, Niazi, Jhanjhi, Alshayeb, and
  Mahmood]{Humayun2020}
\BIBentryALTinterwordspacing
M.~Humayun, M.~Niazi, N.~Z. Jhanjhi, M.~Alshayeb, and S.~Mahmood, ``Cyber
  security threats and vulnerabilities: A systematic mapping study,''
  \emph{Arabian Journal for Science and Engineering}, vol.~45, no.~4, pp.
  3171--3189, Apr 2020. [Online]. Available:
  \url{https://doi.org/10.1007/s13369-019-04319-2}
\BIBentrySTDinterwordspacing

\bibitem[Buro and Mastroeni(2018)]{10.1007/978-3-319-73721-8_6}
S.~Buro and I.~Mastroeni, ``Abstract code injection,'' in \emph{Verification,
  Model Checking, and Abstract Interpretation}, I.~Dillig and J.~Palsberg,
  Eds.\hskip 1em plus 0.5em minus 0.4em\relax Cham: Springer International
  Publishing, 2018, pp. 116--137.

\bibitem[Priyanka and Sai~Smruthi(2020)]{9278437}
A.~K. Priyanka and S.~Sai~Smruthi, ``Web application vulnerabilities:
  Exploitation and prevention,'' in \emph{2020 International Conference on
  Electrotechnical Complexes and Systems (ICOECS)}, 2020, pp. 1--5.

\bibitem[Denis et~al.(2016)Denis, Zena, and Hayajneh]{7494156}
M.~Denis, C.~Zena, and T.~Hayajneh, ``Penetration testing: Concepts, attack
  methods, and defense strategies,'' in \emph{2016 IEEE Long Island Systems,
  Applications and Technology Conference (LISAT)}, 2016, pp. 1--6.

\bibitem[Devi and Kumar(2020)]{9143018}
R.~S. Devi and M.~M. Kumar, ``Testing for security weakness of web applications
  using ethical hacking,'' in \emph{2020 4th International Conference on Trends
  in Electronics and Informatics (ICOEI)(48184)}, 2020, pp. 354--361.

\bibitem[Kumar et~al.(2017)Kumar, Mahajan, Kumar, and Khatri]{8342469}
S.~Kumar, R.~Mahajan, N.~Kumar, and S.~K. Khatri, ``A study on web application
  security and detecting security vulnerabilities,'' in \emph{2017 6th
  International Conference on Reliability, Infocom Technologies and
  Optimization (Trends and Future Directions) (ICRITO)}, 2017, pp. 451--455.

\bibitem[Albalawi et~al.(2023)Albalawi, Alamrani, Aloufi, Albalawi, Aljaedi,
  and Alharbi]{electronics12122664}
\BIBentryALTinterwordspacing
N.~Albalawi, N.~Alamrani, R.~Aloufi, M.~Albalawi, A.~Aljaedi, and A.~R.
  Alharbi, ``The reality of internet infrastructure and services defacement: A
  second look at characterizing web-based vulnerabilities,''
  \emph{Electronics}, vol.~12, no.~12, 2023. [Online]. Available:
  \url{https://www.mdpi.com/2079-9292/12/12/2664}
\BIBentrySTDinterwordspacing

\bibitem[Cremer et~al.(2022)Cremer, Sheehan, Fortmann, Kia, Mullins, Murphy,
  and Materne]{Cremer2022}
\BIBentryALTinterwordspacing
F.~Cremer, B.~Sheehan, M.~Fortmann, A.~N. Kia, M.~Mullins, F.~Murphy, and
  S.~Materne, ``Cyber risk and cybersecurity: a systematic review of data
  availability,'' \emph{The Geneva Papers on Risk and Insurance - Issues and
  Practice}, vol.~47, no.~3, pp. 698--736, Jul 2022. [Online]. Available:
  \url{https://doi.org/10.1057/s41288-022-00266-6}
\BIBentrySTDinterwordspacing

\bibitem[Hehir et~al.(2021)Hehir, Zeller, Luckhurst, and Chandler]{Hehir2021}
\BIBentryALTinterwordspacing
E.~Hehir, M.~Zeller, J.~Luckhurst, and T.~Chandler, ``Developing student
  connectedness under remote learning using digital resources: A systematic
  review,'' \emph{Education and Information Technologies}, vol.~26, no.~5, pp.
  6531--6548, Sep 2021. [Online]. Available:
  \url{https://doi.org/10.1007/s10639-021-10577-1}
\BIBentrySTDinterwordspacing

\bibitem[Challapalli et~al.(2021)Challapalli, Kaushik, Suman, Shivahare, Bibhu,
  and Gupta]{9673464}
S.~S.~N. Challapalli, P.~Kaushik, S.~Suman, B.~D. Shivahare, V.~Bibhu, and
  A.~D. Gupta, ``Web development and performance comparison of web development
  technologies in node.js and python,'' in \emph{2021 International Conference
  on Technological Advancements and Innovations (ICTAI)}, 2021, pp. 303--307.

\bibitem[Palomino et~al.(2021)Palomino, Paz, and
  Moquillaza]{10.1007/978-3-030-78221-4_21}
F.~Palomino, F.~Paz, and A.~Moquillaza, ``Web analytics for user experience: A
  systematic literature review,'' in \emph{Design, User Experience, and
  Usability: UX Research and Design}, M.~M. Soares, E.~Rosenzweig, and
  A.~Marcus, Eds.\hskip 1em plus 0.5em minus 0.4em\relax Cham: Springer
  International Publishing, 2021, pp. 312--326.

\bibitem[Choi and Sim(2016)]{Choi2016}
\BIBentryALTinterwordspacing
H.~Choi and S.~Sim, ``A study on efficiency of markup language using dom
  tree,'' \emph{Wireless Personal Communications}, vol.~86, no.~1, pp.
  143--163, Jan 2016. [Online]. Available:
  \url{https://doi.org/10.1007/s11277-015-3057-z}
\BIBentrySTDinterwordspacing

\bibitem[Rabinovich et~al.(2017)Rabinovich, Stern, and
  Klein]{DBLP:journals/corr/RabinovichSK17}
\BIBentryALTinterwordspacing
M.~Rabinovich, M.~Stern, and D.~Klein, ``Abstract syntax networks for code
  generation and semantic parsing,'' \emph{CoRR}, vol. abs/1704.07535, 2017.
  [Online]. Available: \url{http://arxiv.org/abs/1704.07535}
\BIBentrySTDinterwordspacing

\bibitem[Zheng et~al.(2020)Zheng, Zhao, Long, Zhu, Zhu, Zhao, Diao, Yang, and
  Lin]{DBLP:journals/corr/abs-2009-10924}
\BIBentryALTinterwordspacing
Z.~Zheng, P.~Zhao, G.~Long, F.~Zhu, K.~Zhu, W.~Zhao, L.~Diao, J.~Yang, and
  W.~Lin, ``Fusionstitching: Boosting memory intensive computations for deep
  learning workloads,'' \emph{CoRR}, vol. abs/2009.10924, 2020. [Online].
  Available: \url{https://arxiv.org/abs/2009.10924}
\BIBentrySTDinterwordspacing

\end{thebibliography}
\bibliographystyle{IEEEtranN}

\end{document}